# Predictability of Shanghai Stock Market by Agent-based Mix-game Model

Chengling Gou
Physics Department, Beijing University of Aeronautics and Astronautics
No.37, Xueyuan Road, Heidian District, Beijing, 100083, China
Physics Department, University of Oxford
Clarendon Laboratory, Parks Road, Oxford, OX1 3PU, UK
gouchengling@hotmail.com

*Abstract*—This paper[1] reports the effort of using agent-based mix-game model to predict financial time series. It introduces simple generic algorithm into the prediction methodology, and gives an example of its application to forecasting Shanghai Index. The results show that this prediction methodology is effective and agent-based mix-game model is a potential good model to predict time series of financial markets.

I. INTRODUCTION

Forecasting financial markets is a continuous effort of researchers, and attracts great attention of practitioners. The use of agent-based models for forecasting financial markets is a new attempt in this area. Neil F. Johnson etc. reported on a technique based on multi-agent games which has potential use in predicting future movements of financial time-series. In their articles, a "third-party" game is trained on a "black-box" time-series, and then run into the future to extract next step and multi-step predictions [1, 2, 3]. They used minority game as the third party game. In order to improve the forecasting accurate rate, I use mix-game [13,17] as the third party game. I found that using mix-game can improve forecasting accurate rate at least 3% more than using minority game. I also found the forecasting accurate rate is sensitive to the initial strategy distribution (ISD) of agents. Therefore, I introduce simple generic algorithm into the forecasting process to select the better ISD. This paper reports the results of using agent-based mix-game model to predict financial time series.

A.  *Mix-game Model*

Challet and Zhang's MG model, together with the original bar model of Arthur, attracts a lot of following studies. Given the MG's richness and yet underlying simplicity, the MG has also received much attention as a financial market model [4]. The MG comprises an odd number of agents choosing repeatedly between the options of buying (1) and selling (0) a quantity of a risky asset. The agents continually try to make the minority decision, i.e. buy assets when the majority of other agents are selling, and sell when the majority of other agents are buying. Neil Johnson [5, 6] and coworkers extended MG by allowing a variable number of active traders at each timestep--- they called their modified game as the Grand Canonical Minority Game (GCMG). The GCMG, and to a lesser extent the basic MG itself, can reproduce the stylized facts of financial markets, such as volatility clustering and fat-tail distributions.

However, there are some weaknesses in MG and GCMG. First, the diversity of agents is limited since agents all have the same memory length and time-horizon. Second, in real markets, some agents are tendency-followers, i.e. "noise traders" [7~12, 18~22], who effectively play a majority game; while others are "foundation traders", who effectively play a minority game.

In order to create an agent-based model which more closely mimics a real financial market, C. Gou modifies the MG model by dividing agents into two groups: each group has different memory and time-horizon, and one group play the minority game and the other play the majority game. For this reason, this system is referred to as a 'mix-game' model [13, 17].

Since mix-game model is an extension of minority game (MG), its structure is similar to MG. In mix-game, there are two groups of agents; group1 plays the majority game, and group2 plays the minority game. N (odd number) is the total number of the agents, and N1 is number of agents in group1. The system resource is R=N*L, where L<1 is the proportion of resource of the system. All agents compete in the system for the limited resource R. T1 and T2 are the time horizon lengths of the two groups, and m1 and m2 denote the memory lengths of the two groups, respectively.

The global information only available to agents is a common bit-string "memory" of the m1 or m2 most recent competition outcomes (1 or 0). A strategy consists of a response, i.e. 0 (sell) or 1 (buy), to each possible bit string; hence there are $2^{2^{m1}}$ or $2^{2^{m2}}$ possible strategies for group1 or group2, respectively, which form full strategy spaces (FSS). At the beginning of the game, each agent is assigned s strategies and keeps them unchangeable during the game. After each turn, agents assign one (virtual) point to a strategy which would have predicted the correct outcome. For agents in group1, they reward their strategies one point if they are in the majority; for agents in group2, they reward their strategies one point if they are in the minority. Agents collect the virtual points for their strategies over the time horizon T1 or T2, and they use their strategy which has the highest virtual point in each turn. If there are two strategies which have the highest virtual point, agents use coin toss to decide which strategy to be used.

---
[1] This research was performed in collaboration with Professor Neil F. Johnson, Oxford University.

Excess demand is equal to the number of ones (buy) which agents choose minus the number of zeros (sell) which agents choose. According to a widely accepted assumption that excess demand exerts a force on the price of the asset and the change of price is proportion to the excess demand in a financial market, the time series of the price can be calculated based on the time series of the excess demand [14, 15, 16].

Through simulations, C. Gou finds out that the fluctuations of local volatilities change greatly by adding some agents who play majority game into MG, but the stylized features of MG don't change obviously except agents with memory length 1 and 2, and the correlation between the average winning of agents and the mean of local volatilities of the systems is different with different configurations of m1 and m2. She also gives suggestions about how to use mix-game model to simulate financial markets, and shows the example of modeling Shanghai stock market by means of mix-game model [13, 17].

*B. Prediction Method*

We use mix-game model to do one-step direction prediction about a real time series produced by a financial market, whose dynamics are well-described by a mix-game model for a unknown parameter configuration of T1, T2, m1, m2, N, N1 and an unknown specific realization of initial strategy distributions (ISD). We call this as the "black-box" game. Based on mix-game model, we need to identify "third-party" game which can be matched with the black-box game. After we select the parameter configuration for the third party game, we use it to predict the time series produced by the black-box game [1~3]. Fig.1 shows the work flow for prediction.

In the third party game, agents receive the global information which is strings of 1's and 0's (WinEx) digitized from the real time series produced by the black box game other than that (WinIn) generated by the third party game itself, and agents also reward their strategies according to the real time series (WinEx). In order to see the performances of the third party game, we compare the time series generated by the third party game with the real time series produced by the black box game, i.e. we compare WinIn with WinEx because we just predict the direction. If they match, the third party game gets one point; if not, it gets zero. This point is referred to as accurate scores. At each predicting timesteps, we calculate the accurate rate which is equal to accurate scores divided by predicting timesteps.

Accurate rate (current time)
=accurate scores / predicting timesteps          (1)

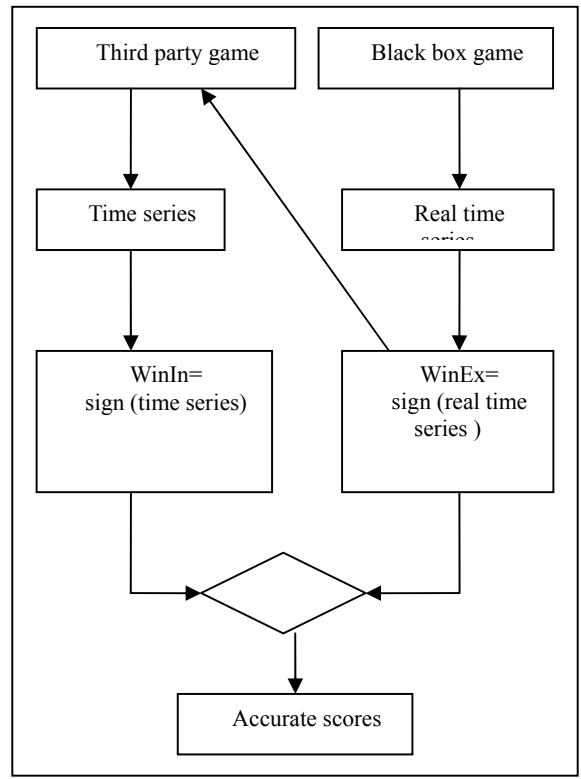

Fig.1 work flow for prediction

*C. Forecasting Process*

*1) Searching for Parameters of Mix-game Model to Simulate a Specific Financial Market*

First we need to search for proper parameter configurations for modeling the target financial market in order to identify the third party game. According to [13, 17], the mix-game model with some parameter configurations reproduces the stylized features of financial time series.

Therefore, we need to choose parameters of m1, m2, T1, T2, N and N1 when using mix-game to model financial markets. To do so, the following aspects need to be considered.

- First make sure the time series of mix-game can reproduce the stylized facts of time series of financial markets by choosing proper parameters, especially m1 and m2 must be larger than 2, and m1< m2;
- Second pay attention to the fluctuation of local volatilities, and ensure that the median of local volatilities of mix-game is similar to that of the target time series;
- Third make sure the log-log plot of absolute returns look similar.

*2) Searching for the Better ISD by Using Simple Generic Algorithm*

Second, the ISD need to be chosen in full strategy space (FSS) in order to improve accurate rates because the accurate rates are greatly sensitive to ISD. However, the FSS is huge and can not be searched thoroughly by trial-error method. In order to perform the search, I use simple generic algorithm (SGA) with different parameter configurations of mix-game model for the target financial market.

Since I use a set of parameters to generate ISD, the problem of searching for the better ISD is a problem of searching for a set of parameters of ISD. In order to use SGA to optimize parameter sets, first I code the parameter set into a set of strings of zeros and ones, which are "chromosomes"; then I attach fitness to them. In our case, the fitness is just the accurate rate. The following is SGA where P(t) indicates the population of parameter sets used to generate ISD.

```
SGA( )
%{   t=0;
%    initialize P(t);
     %   evaluation P(t);
%}
% while (not terminal-condition )
% {
%    t=t+1;
           %   select P(t) from P(t-1);
           %   recombined P(t);
           %   evaluation P(t);
  %}
```

At this stage, usually one needs to adjust N1, T1 and T2 so as to get better ISD. Generally speaking, the parameters of N1, T1 and T2 for forecasting are slightly different from that for simulating a financial market

## II. MAIN RESULTS AND DISCUSSION

As an example, I use this methodology to forecast Shanghai Index dating from 02-07-2002 to 30-12-2003

### A. Mix-game Model for Shanghai Stock Market

Since the median value of local volatilities of Shanghai Index daily data from 1992/01/02 to 2004/03/19 is 222, two parameter configurations of mix-game model have the similar median values of local volatilities according to [13]; one configuration is m1=3, m2=6, T1=12, T2=60, N=201, N1=72, the other is m1=4, m2=6, T1=T2=12, N=201, N1=72[13].

### B. Optimal Parameters and Their Forecasting Accurate Rates

TABLE I

OPTIMAL PARAMETER CONFIGURATIONS AND THEIR FORECASTING ACCURATE RATES AT 200 TIMESTEPS AS WELL AS ACCURATE RATES OF CORRESPONDING MG (I.E. N1=0)

| N0 | Parameter configurations of mix-game | Accurate rates of mix-game | Accurate rates of MG |
|---|---|---|---|
| 1 | m1=3, m2=6, N=201, N1=90, T1=24, T2=24, | 61.3% | 45.7% |
| 2 | m1=3, m2=6, N=201, N1=90, T1=12, T2=24, | 60.3% | 49.3% |
| 3 | m1=3, m2=6, N=201, N1=95, T2=12, T1=12, | 61.8% | 51.8% |
| 4 | m1=4, m2=6, N=201, N1=95, T1=12,T2=24, | 59.3% | 49.3% |
| 5 | m1=4, m2=6, N=201, N1=90, T1=24, T2=24, | 61.3% | 49.8% |
| 6 | m1=4, m2=6, N=201, N1=85, T1=12, T2=24, | 60.8% | 48.7% |
| 7 | m1=4, m2=6, N=201, N1=80, T1=18, T2=24, | 60.8% | 48.7% |

Table 1 lists the resulting optimal parameter configurations and their forecasting accurate rates at 200 timesteps as well as accurate rates of corresponding MG (i.e. N1=0) related to different specific ISD found by SGA program. Most accurate rates for mix-game are larger than 60% while they are at least 10% smaller for MG with the same parameters and ISD. This results show that mix-game model really can improve the prediction accurate rates.

Fig.2. shows an example of the in sample time series of the accurate rates of the parameter configuration of m1=3, m2=6, T1=T2=24, N1=90, N=201 with a specific ISD. One

can find that the accurate rate has high value at short time period which is around 30 timesteps and the accurate rate reaches a stable level from 120 timesteps to 200 timesteps. The time series of the accurate rates of other parameter configurations behave similar to Fig.2.

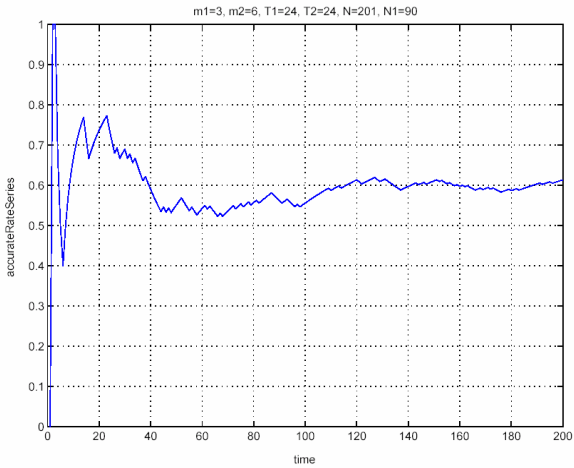

Fig.2. In sample time series of the accurate rates of parameter configuration of m1=3, m2=6, T1=T2=24, N1=90, N=201 with a specific ISD

However, we notice that the landscape of the accurate rates vs. the parameter sets of ISD seems rugged. So the better accurate rates are just obtained by the present SGA program, and there are chances to improve SGA program so as to get better accurate rates.

## C. Out of Sample Test

I do out of sample test by shifting the forecasting time window which is 200 timesteps. First, I shift the time window 100 timesteps (T3=100) forward along the time series of Shanghai Index. Second, we shift the time window 215 timesteps (T3=215) forward along the time series of Shanghai Index.

Fig.3 shows the resulting time series of the accurate rates of parameter configuration of m1=3, m2=6, T1=T2=24, N1=90, N=201, T3=100 with the same specific ISD as that in sample. From Fig.3, we can find that the accurate rate is higher than 60% after forecasting timestep is 40. The time series of accurate rates of other parameter configurations behave similar to Fig.3.

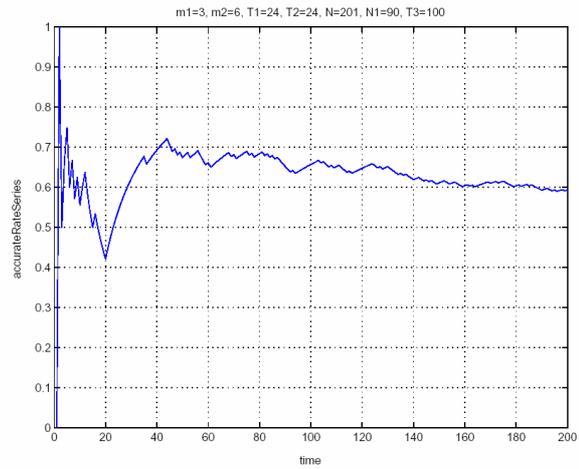

Fig.3. out of sample time series of the accurate rates of parameter configuration of m1=3, m2=6, T1=T2=24, N1=90, N=201, T3=100 with the same specific ISD as that in sample

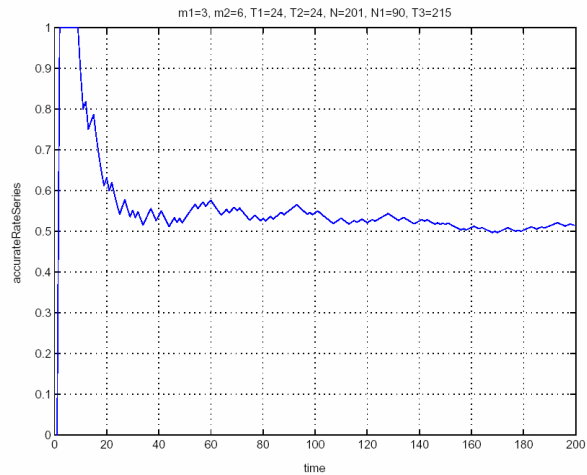

Fig.4 time series of the accurate rates of parameter configuration of m1=3, m2=6, T1=T2=24, N1=90, N=20, T3=215 with the same specific ISD as that in sample

Fig.4. shows the resulting time series of the accurate rates of parameter configuration of m1=3, m2=6, T1=T2=24, N1=90, N=201, T3=215 with the same specific ISD as that in sample. From Fig.4, we can see that the accurate rate drops quickly and only a little above 50% after forecasting timestep is 40. The time series of accurate rates of other parameter configurations behave similar to Fig.4.

The results of out of sample test suggest that we need to tune ISD along the time series which we try to forecast. We will report the study on this issue elsewhere [23].

## III. CONCLUSIONS

In conclusion, our prediction methodology is effective and agent-based mix-game model is a potential good model to predict time series of financial markets. The further study will focus on the improvement of SGA program in order to find better ISD so as to get better accurate rates. And another issue which is worthwhile to study is how the ISD changes along the time series which we try to forecast.

## ACKNOWLEDGMENT

This research is supported by China Scholarship Council. Thanks Professor Yaping Lin for helpful discussion..

## REFERENCES

[1] Neil F. Johnson,, David Lamper, Paul Jefferies Michael L. Hart and Sam Howison, Application of multi-agent games to the prediction of financial time-series, Physica A 299 (2001) 222–227.
[2] D. Lamper, S. D. Howison and N. F. Johnson, Predictability of Large Future Changes in a Competitive Evolving Population, Physical Review Letters Vol. 88, No.1, 017902-1~4 (2002).
[3] P. Jefferies, M.L. Hart, P.M. Hui, and N.F. Johnson, From market games to real-world markets, Eur. Phys. J. B 20, 493{501 (2001).
[4] W.B. Arthur, Science 284, 107 (1999); D. Challet, and Y. C. Zhang, Phyisca A 246, 407(1997);
[5] Neil F. Johnson, Paul Jefferies, and Pak Ming Hui, Financial Market Complexity, Oxford Press(2003);
[6] Paul Jefferies and Neil F. Johnson, Designing agent-based market models, Oxford Center for Computational Finance working paper: OCCF/010702;
[7] J. V. Andersen, and D. Sornette, The $-game, cond-mat/0205423;
[8] Challet, Damien, Inter-pattern speculation: beyond minority, majority and $-games, arXiv: physics/0502140 v1.
[9] F. Slanina and Y.-C. Zhang, Physica A 289, 290 (2001)
[10] T. Lux, Herd Behavior, Bubbles and Crashes. Economic Journal 105(431): 881-896(1995). T. Lux, and M. Marchesi Scaling and criticality in a stochastic multi-agent model of a financial market. Nature 397(6719): 498-500 (1999).
[11] J.V. Andersen and D. Sornette, Eur. Phys. J. B 31, 141 (2003)
[12] I. Giardina and J.-P. Bouchaud, Eur. Phys. J. B 31, 421 (2003)
[13] Chengling Gou, Dynamic Behaviors of Mix-game Models and Its Application, /arxiv.org/abs/physics/0504001
[14] J.P. Bouchaud and R. Cont, Eur. Phys. J. B 6 543 (1998)
[15] J. D. Farmer, adap-org/9812005.
[16] J. Lakonishok, A. Shleifer, R. Thaler and R. Vishny, J. Fin. Econ. 32, 23 (1991)
[17] Chengling Gou, Agents Play Mix-game, http://arxiv.org/abs/physics/0505112
[18] M. Marsili, Physica A 299 (2001) 93 .
[19] A. De Martino et al, J. Phys. A: Math. Gen. 36 (2003) 8935
[20] A C C Coolen, The Mathematical Theory of Minority Games, Oxford University Press, Oxford(2005).
[21] A. Tedeschi, A. De Martino and I. Giardina, http://arXiv.org/abs/cond-mat/0503762 .
[22] A. De Martino, I. Giardina1, M. Marsili and A. Tedeschi, http://arXiv.org/abs/cond-mat/0403649
[23] Chengling Gou, Neil F. Johnson ( in preparation).